\documentstyle[12pt]{article}
\begin{document}
\begin{titlepage}

\begin{flushright}
REVISED
\end{flushright}

\begin{center}
{\bf {\LARGE Geometric Bremsstrahlung \\
in the Early Universe }} \\

\vspace*{2cm}
{\Large T.\ Futamase}\\
{\it Astronomical Institute, Faculty of Science,\\ Tohoku University, 
Sendai 980-77, Japan}\\

\vspace*{1cm}
{\Large M.\ Hotta,\ \ \    H.\ Inoue\ \   }\\
{\it Department of Physics, Faculty of Science, \\ Tohoku University, 
Sendai 980-77, Japan}\\

\vspace*{1cm}
{\Large   M.\ Yamaguchi}\\
{\it Department of Physics, Faculty of Science, \\ Tohoku University, 
Sendai 980-77, Japan.\\
Institute f\"{u}r Theoretical Physik, Physik Department,\\ 
Technische Universit\"{a}t M\"{u}nchen 85747 Garching, Germany}\\

\end{center}

\vspace*{1cm}

\begin{abstract}
 We discuss photon emission from particles  
 decelerlated by the cosmic expansion. This can be interpretated  
as a kind of bremsstrahlung induced by the Universe geometry.
 In the high momentum limit  its transition probability
 does not depend on detailed behavior
 of the expansion. This may play an important role when 
massless particle emission is discussed in the early Universe.
\end{abstract}

\end{titlepage}

\section{Introduction}

\hspace{0.8cm} Particles in the expanding Robertson-Walker 
Universe are decelerated in the comoving frame.
Ignoring backreaction, they obey the geodesic equation and lose
 their physical momentum $P_{phys}$ as
\begin{equation}
P_{phys} = \frac{P_{conf}}{a} \rightarrow 0, 
\end{equation} 
where $P_{conf}$ is a conserved conformal momentum and $a$ is 
a scale factor growing in time. In general 
 particles in deceleration can
 emanate  radiation, or some massless particles.
 We call this process  geometric bremsstrahlung\footnote{ This process 
is also called by DeWitt and Brehme   
electro-gravitic bremsstrahlung in ref \cite{DB}.} due to
 the cosmic expansion.

 Many analyses on phenomena in 
the early Universe have been performed so far
using results of high energy particle physics and proposed
 a lot of interesting features of the Universe\cite{KT}. 
However they are based on calculations of transition matrices  
in the flat spacetime, emphasizing the fact that rates 
of interactions  are much larger than expansion rate given by 
 Hubble parameter,
and no careful attention seems to be paid to the geometric bremsstrahlung
process, which actually  gives no contribution in the flat spacetime.

 We shall argue 
 in this paper that geometric bremsstrahlung may play
impotant roles in the early Universe. In the section 2, emission of
electromagnetic wave
from a classical charged particle in the expanding Universe  
 is discussed, taking account of backreaction. 
Our argument in the classical level suggests that the  
emission rate may not be simply ignored and the damping time
 may nearly equal to expansion  time.
 Thus in the section 3, we treat photon emission from charged a particle
 quantum mechanically. 
High momentum limit of the transition probability can be
obtained analytically. We stress that massless limit 
 should be treated carefully and is nontrivial.

In this paper, we adopt the natural units, the light velocity $c=1$ and 
the Planck constant $\hbar =1$. Signature of metric is taken as $(+,-,-,-)$.

\section{Classical Geometric Bremsstrahlung in the Early Universe}

\hspace{0.8cm}The radiation reaction has 
been neglected in the study of the early
Universe. However particles are deaccelerated due to the cosmic
expansion and thus they  will  emit the radiation. If the damping time
due to the radiation reaction is comparable to the expansion time, 
the effect of the radiation reaction may not be simply neglected. 
We shall study in detail this phenomenon in the case of 
classical charged particles.

The study of the radiation reaction for a charged particle has a long 
history. The first relativistic calculation was performed by 
Dirac\cite{Dirac}.
His calculation has been generalized by  DeWitt-Brehme\cite{DB}
for the motion 
in gravitational field. They have shown that bremsstrahung induced by
the spacetime curvature which we call geometric bremsstrahlung
 occured in addition to the usual radiation
damping. The effect is nonlocal in general which is caused by the
so-called tail term in the Green function. 
It was  Hobbs\cite{Hobbs} 
who corrected the result of De Witt-Brehme and pointed
out that the tail term vanishes identically in the case of the 
conformally flat spacetimes.
  His equation of motion for a particle with 4 velocity  
$ u^\mu $, mass $m$, charge $e$ without  external electromagnetic
field  may be written in the following form in conformally flat spacetime.
\begin{eqnarray} 
m \frac{D u^\mu}{D\tau} 
=  \frac{2 e^2}{3} 
\left( 
\frac{D^2 u^\mu}{D \tau^2} 
 + u^\mu
 \left( 
\frac{D u}{D\tau} 
\right)^2 
\right)   
+\frac{2e^2}{3}  
( \Omega_{,\alpha\beta} 
  - \Omega_{,\alpha} \Omega_{,\beta} ) 
  \left(  
    g^{\mu\alpha} u^\beta 
       - u^\mu u^\alpha u^\beta 
  \right)  \label{20}
\end{eqnarray}
where $D/D\tau $ is the absolute derivative along the worldline of the 
particle with $ \tau$  the  proper time and the $ \exp (2 \Omega ) $ is 
the conformal factor, $ g_{\mu\nu} = e^{2\Omega} \eta_{\mu\nu} $
with $ \eta_{\mu\nu} $  the flat Minkowskii metric. 

Here we are  interested in the 
radiation reaction induced by the cosmic expansion in the early
Universe and thus we will  restrict ourselves to the case where 
the conformal factor depends only on the time variable. 
We shall take a different approach from Hobbs and
evaluate explicitly the damping time scale due to the
geometric bremsstrahlung
 in the
early Universe. 

\bigskip

We shall take the standard form for the action of a charged particle 
with mass $m$ and charge $e$ in the gravitational field, 

$$  S = - m \int \sqrt{g_{\mu\nu} {\dot{x}}^\mu 
                                 {\dot{x}}^\nu } d\tau 
        - e \int A_\mu \dot{x}^\mu d\tau 
        - {1\over 4} \int d^4 x \sqrt{-g} 
          g^{\alpha\beta} g^{\mu\nu} F_{\alpha\mu} F_{\beta\nu} . 
$$
The dot denotes the derivative with respect to the proper time $\tau$.
The equation of motion derived from the above action may be written as 
follows.
\begin{eqnarray}
 m {{D u^\mu}\over {D\tau}} 
   =  m \left( {{du^\mu}\over {d\tau}} 
        + \Gamma^\mu_{\alpha\beta}u^\alpha u^\beta \right)
   = e F^{\mu\nu} u_\nu.\nonumber
\end{eqnarray}

Since we are interested in the early Universe, we may neglect the
spatial curvature and thus take the spatially flat Robertson-Walker 
model as our background geometry,

$$ ds^2 = a^2 (\eta) ( d\eta^2 - d{\vec{x}}^{\ 2} )
 =dt^2 -a^2 d{\vec{x}}^{\ 2} .
$$

Since the metric is conformally flat, it is convenient to work 
in the conformally related flat spacetime.
Defining the conformally related proper time $ d\tau_f = a^{-1}
d\tau $, we shall define the conformally related  4 velocity

$$ {\tilde u}^\mu  = {{dx^\mu}\over {d\tau_f}} . $$

Then the equation of motion  may be written as follows.

$$ m {{D  u^\mu}\over {D\tau}} 
    = m a^{-2}   \left( {{d{\tilde u}^\mu}\over {d\tau_f}} 
    + {{a'}\over a} {{d\eta}\over {d\tau_f}} {\tilde u}^\mu 
    - {{a'}\over a} \delta^\mu_0  \right) 
   = e  a^{-3} \eta^{\mu\alpha} F_{\alpha\beta} {\tilde u}^\beta .
$$
where the prime denotes the derivative with respect to the conformal 
time $\eta $. By using the self field of the particle in the right 
hand side of the above equation, we shall obtain the radiation
reaction force. 

Before calculating the reaction force explicitely, let us compare 
the timescale due to the radiation damping with that due to
 the cosmic expansion 
to see the importance of the radiation reaction in the early
Universe. The damping time may be roughly evaluated as follows.

$$ {1\over {t_{r}}} \sim 
 \left| {1\over E_{conf}} {{dE_{conf}}\over {ad\eta}} \right| 
         = {1\over p_{conf}} \, {2\over 3 a} e^2  
\left({{d{\tilde{u}} }\over {d\eta}}\right)^2 
 =\frac{2e^2}{3p_{conf} a}
\left(\frac{p_{conf} }{m a^2} \frac{da}{d\eta} \right)^2 
    = {2\over 3} e^2 \, {{p_{phys} H^2}\over {m^2}}   
$$ 
where 
$$
H= \frac{1}{a}\frac{da}{dt} 
$$
 is the Hubble parameter, $p_{conf}$ is the conformal momentum
and 
we have used the
fact that the physical momentum $ p_{phys}= m{\tilde{u}}=p_{conf}/a $ 
decays as $ a^{-1}$ if the
radiation reaction is neglected. 
Thus the ratio between the Hubble time $ t_{exp} = H^{-1} $ and 
the damping time is 
$$  {{t_{exp}}\over {t_{r}}} 
        \sim  {{2e^2}\over {3m^2}} p_{phys} H    . 
$$
This ratio is much larger than the unity for a relativistic
particle at sufficiently early times in the Universe.  
Thus  the radiation reaction may not be simply ignored and might play 
an important role in the early Universe. 

\bigskip

For the calculation of the reaction force, we shall need 
the  field equation derived from the above action,

$$ \eta^{\mu\nu} \eta^{\alpha\beta} F_{\alpha\nu,\beta} 
   = e \int d \tau_f \delta^4 ( x - x( \tau_f)) {\tilde u}^\mu .  
$$
Taking the following non-covariant gauge

$$ \eta^{\mu\nu} A_{\mu,\nu} = 0,     $$
we arrive at the field equation which has the same form with that in 
the flat spacetime,
$$  \eta^{\alpha\beta} A^\mu_{,\alpha\beta} 
     = e \int d \tau_f {\tilde u}^\mu \delta^4 ( x - x(\tau_f))  .
$$ 
Then the calculation by Dirac \cite{Dirac} applies here and we  obtain  
the standard expression for the
reaction force in the flat spacetime,
$$  F^{\mu}_{react} =e\  \eta^{\mu\alpha} F_{\alpha\beta} {\tilde u}^\beta 
      =  {2\over 3} e^2  
         \left[ {{d^2 {\tilde u}^\mu}\over {d\tau_f}^2}
         + \left( {{d{\tilde u}}\over {d\tau_f}}\right)^2 
           {\tilde u}^\mu \right] .
$$
It can be shown by a direct calculation that our expression of the
equation of motion with radiation reaction in the conformally related
flat spacetime coincides with eqn(\ref{20})  
when transformed back to the original physical frame.

\bigskip

In order to see the effect of the radiation reaction explicitly, 
we shall forcus our attention to an  1-dimensional motion. Then the 
above equation is simplified as 
$$ {d\over {d\tau_f}} ( a m {\tilde u} ) 
    = {2\over 3} e^2  \left[ {{d^2{\tilde u}}\over {d\tau_f}^2} 
       - {{\tilde u}\over { 1 + {\tilde u}^2}}  
         \left({{d{\tilde u}}\over {d\tau_f}}\right)^2  \right] .
$$
Without the radiation reaction, the conformal momentum 
$ p_{conf} = am{\tilde{u}} $ is conserved as expected. 

Now we shall rewrite the above equation using  the conformal 
momentum $p_{conf}$ and the background time $ dt = a d\eta $, 
\begin{eqnarray}
 {{d^2 p_{conf}}\over {dt^2}} 
    = \left( H + {{3m}\over {2e^2 \sqrt{ 1 + (p_{phys}/m)^2}}} \right) 
      {{d p_{conf}}\over {dt}} + {{dH}\over {dt}} p_{conf}    .
\label{21}
\end{eqnarray}
Notice that there will be no classical geomotric bremsstrahlung in the case 
of de Sitter expansion, namely $H= const$. 
We shall be interested in the  relativistic case in the early
universe, namely 

$$  p_{phys} \gg {{3m^2}\over {2e^2 H}}, \; \; \; m  . $$
Then the second term in the coefficient of $ dp_{conf}/dt $ 
in eqn(\ref{21})  
is negligible. 
Thus when the particle is  relativistic, its evolution is governed 
by the reaction force only and the Hubble time will be  the only 
available time scale in this situation. 
In fact, the solution in this case may be written as follows.

$$ p_{conf} (t) = p_0 \left( 1 - H(t_0) \int^t_{t_0} d t' 
             \exp \left( - \int^{t'}_{t_0} H(x) dx \right) \right) 
             \exp \left( \int^t_{t_0} H(t') dt' \right)  
$$
where we have taken the following initial conditions;

$$ p_{conf} (t=t_0) = p_0, \; \; \; {{dp_{conf}}\over {dt}}(t=t_0) = 0 $$
The second condition expresses the fact that the reaction force is
absent at the initial time. The solution shows that the momentum decays 
in the Hubble time. Thus this process should not be simply neglected. 
 However the
above conclusion is obtained  as a classical effect and it is not clear
if the geometric bremsstrahlung is still effective when the quantum effect 
 is taken into account. We shall discuss quantum geometrobremsstrahlung 
in the next section.

\section{Quantum Geometric Bremsstrahlung}

\hspace{0.8cm}  As argued in the section 2, 
 geometric bremsstrahlung may work efficiently  classical mechanically 
in the early Universe.  Whether
 this notable process survives or not after 
taking quantum effects into account is rather nontrivial and this  question
will be addressed next.

 To define well-behaved quantum amplitudes in the expanding Universe,
 we consider spacetimes with Minkowskian in- and out- regions. The way of 
expansion is chosen arbitrary. The scale factor is described as 
\begin{equation}
a(\eta) = C(\lambda \eta)\label{1001}
\end{equation}
where $\eta$ is the conformal time, $\lambda^{-1}$ is a constant exhibiting 
typical time scale of the expansion.
The function  $C(x)$ in eqn(\ref{1001}) is arbitrary except the following
constraints,
\begin{eqnarray}
C(x\sim -\infty ) &=&b ,\label{1011}\\
C(x\sim \infty) &=&1,\label{1012}\\
C(x) &>& 0,\label{1013}
\end{eqnarray}
where $b$ is some positive constant describing ratio of initial scale
factor to final one.

Consider first the photon emission process 
 in massive scalar QED with conformal coupling to the background 
curvature. 
The action reads 
\begin{eqnarray}
S= \int d^4 x &\sqrt{-g}
&\left(
(\nabla_{\mu} +ieA_{\mu}) \Phi^{\ast} 
(\nabla^{\mu} -ieA^{\mu} )\Phi 
\right. \nonumber\\
&&\left. +(\frac{1}{6} R-m^2 )\Phi^{\ast} \Phi
-  \frac{1}{4} F^{\mu\nu} F_{\mu\nu} 
\right).\nonumber
\end{eqnarray}
The photon emission process is prohibited in the flat spacetime 
by energy-momentum conservation. However in the 
expanding spacetimes the energy
conservation law gets broken and the transition can take place.
  The transition amplitude is given in the lowest order of 
pertubation such that
\begin{equation}
Amp =  -ie 
\int d^4 x \sqrt{-g} 
i\left(
\Phi_f^{\ast}\nabla^{\mu} \Phi_i
-\nabla^{\mu} \Phi_f^{\ast} \Phi_i 
\right) 
A^{\ast}_{\mu} \label{24}
\end{equation}
where $\Phi_i$($\Phi_f$) is initial(final) mode function of
massive charged scalar field and $A^{\ast}_{\mu}$ is final mode
function of electromagnetic field. The scalar mode functions satisfy
\begin{equation}
\left(\nabla^2 +m^2 -\frac{1}{6}R\right) \Phi =0.\nonumber
\end{equation}
Redefining the field $\tilde{\Phi}=\Phi\cdot a(\eta)$, the wave equation
becomes the Klein-Gordon equation with a time-dependent mass,
\begin{equation}
\left(\partial^2 +m^2 a(\eta)^2 \right) \tilde{\Phi} =0.\label{1002}
\end{equation}
Here we introduce $g_{\vec{p}_i}^{in} (\eta)$ and $g_{\vec{p}_f}^{out} (\eta)$
 satisfying
a Schr{\"o}dinger-type equation,
\begin{equation}
\left[-\frac{d^2}{d\eta^2} -m^2 a(\eta)^2 \right]g_{\vec{p}}
=\vec{p}^{\ 2} g_{\vec{p}},\label{9}
\end{equation}
with the boundary conditions in the asymptotic in and out regions as
\begin{eqnarray}
g_{\vec{p}_i}^{in} (\eta) &\rightarrow & 
\frac{
\exp
\left(
-i\eta \sqrt{\vec{p}_i^{\ 2} + m^2 b^2}  
\right)
}{\sqrt{ (2\pi)^3 2 \sqrt{\vec{p}_i^{\ 2} +m^2 b^2}}
}
\ \ \ (\eta\sim -\infty),\nonumber\\
g_{\vec{p}_f}^{out} (\eta) &\rightarrow &\frac{\exp\left(
-i\eta\sqrt{\vec{p}_f^{\ 2} +m^2}
\right)
}{
\sqrt{(2\pi)^3 2\sqrt{\vec{p}_f^{\ 2} +m^2 }}
}
\ \ \ (\eta\sim \infty).\nonumber
\end{eqnarray}
 They also satisfy the following normalization condition. 
\begin{equation}
i\left(g^{\ast}_{\vec{p}} {g'}_{\vec{p}}
-{g'}^{\ast}_{\vec{p}} g_{\vec{p}} \right)=\frac{1}{(2\pi)^3 },\nonumber
\end{equation}
where the prime denotes the derivative with respect to $\eta$.
Then the mode functions can be expressed as
\begin{eqnarray}
\Phi_f^{\ast} &=&\frac{1}{a}\tilde{\Phi}^{\ast}_f
=\frac{1}{a} e^{-i\vec{p}_f \cdot \vec{x} }g^{out\ast}_{\vec{p}_f},\nonumber\\ 
\Phi_i &=&\frac{1}{a}\tilde{\Phi}_i
=\frac{1}{a} e^{i\vec{p}_i \cdot \vec{x} }g_{\vec{p}_I}^{in}.\nonumber
\end{eqnarray}

The electromagnetic final mode function satisfies the Maxwell equation
in curved spacetime,
\begin{equation}
\nabla^{\mu}( \nabla_{\mu} A^{\ast}_{\nu} -\nabla_{\nu} A^{\ast}_{\mu} )
=0.\label{23}
\end{equation}
Notice that in 4-dimensional conformally flat spacetimes
 this eqn(\ref{23}) can be reduced into the
same form  in the flat spacetime,
\begin{equation}
\partial^{\mu}( \partial_{\mu} A^{\ast}_{\nu} 
-\partial_{\nu} A^{\ast}_{\mu} ) =0.\nonumber
\end{equation}
Therefore we get easily the final mode function 
\begin{equation}
A^{\ast}_{\mu} = \epsilon^{\ast}_{\mu} (\vec{k}) 
\frac{\exp\left(
i|\vec{k} |\eta-i \vec{k}\cdot \vec{x}
\right)
}{
\sqrt{(2\pi)^3 2|\vec{k}|}
},\nonumber
\end{equation}
where $\epsilon^{\ast}_{\mu}$ is a helicity factor.

Using the rescaled field, the amplitude, eqn(\ref{24}), is rewritten as
\begin{equation}
Amp =  -ie 
\int d^4 x  
i\left(
\tilde{\Phi}_f^{\ast}\partial^{\mu} \tilde{\Phi}_i
-\partial^{\mu} \tilde{\Phi}_f^{\ast} \tilde{\Phi}_i 
\right) 
A^{\ast}_{\mu}.\nonumber
\end{equation}
Because the photon emission lasts only during the epoch of expansion,
 the concept of the probability {\it per unit time} is ambiguious. So  
we shall use the transition probability itself. 
The transition probability  $W$ 
can be obtained from the amplitude such that
\begin{equation}
W = \sum_{h=L, R} \frac{(2\pi)^3}{V}\int d^3 p_f d^3 k |Amp|^2,\nonumber
\end{equation}
where the summation is performed over the photon helicity and $V$ is
  the conformal volume of the space, which is cancelled by the factor
$(2\pi)^3 \delta(\vec{0})$ 
coming from the conformal momentum conservation in $|Amp|^2$ .
After the helicity summation, the explicit form of $W$ is obtained as 
follows.
\begin{eqnarray}
W &=&
(2\pi)^6 e^2 \int \frac{d^3 k}{(2\pi)^3 2|\vec{k}|} \int d^3p_f 
\delta(\vec{k}+\vec{p}_f -\vec{p}_i )  \nonumber\\
&&\times\left[ (\vec{p}_f +\vec{p}_i )^2 \left| \int
 d\eta \ e^{i|\vec{k}|\eta} 
g^{out\ast}_{\vec{p}_f} g^{in}_{\vec{p}_i} \right|^2
-\left| \int d\eta\ e^{i|\vec{k}|\eta} (
g^{out\ast}_{\vec{p}_f} {g^{in}_{\vec{p}_i}}' 
-{g^{out\ast}_{\vec{p}_f}}' g^{in}_{\vec{p}_i} )
\right|^2
\right]\ .\nonumber
\end{eqnarray}
By virtue of the Wronskian relation;
\begin{equation}
\frac{d}{d\eta}\left(
g^{out\ast}_{\vec{p}_f} {g^{in}_{\vec{p}_i}}' 
-{g^{out\ast}_{\vec{p}_f}}' g^{in}_{\vec{p}_i} 
\right)
=
(\vec{p}_f^{\ 2} -\vec{p}_i^{\ 2} ) g^{out\ast}_{\vec{p}_f} g^{in}_{\vec{p}_i},
\nonumber
\end{equation}
the form of $W$ is more simplified such that
\begin{eqnarray}
W= (2\pi)^3 e^2 \int \frac{d^3 k}{ 2|\vec{k}|}
4\left(
 \vec{p_i}^2 -\frac{ (\vec{k}\vec{p}_i )^2 }{\vec{k}^2 } 
\right)
\left| 
\int^{\infty}_{-\infty} 
d\eta \ e^{i|\vec{k}|\eta} 
g^{out\ast}_{\ \vec{p}_i -\vec{k}} g^{in}_{\vec{p}_i} 
\right|^2 .\label{100}
\end{eqnarray}

To grasp the behavior of $W$ in the $|\vec{p}_i|\rightarrow \infty$ limit, 
we first argue a case with scale factor
\begin{equation}
a(\eta) = \Theta (\eta) + b \Theta (-\eta).\label{8}
\end{equation}
Then the exact mode functions are derived  as
\begin{eqnarray}
g^{in}_{\vec{p}_i} &=& 
\Theta(-\eta)
\frac{
\exp
\left(
-i\eta \sqrt{\vec{p}_i^{\ 2} + m^2 b^2}  
\right)
}{\sqrt{ (2\pi)^3 2 \sqrt{\vec{p}_i^{\ 2} +m^2 b^2}}
}
\nonumber\\
&&+\Theta(\eta)
\frac{
A(\vec{p}_i)\exp
\left(
-i\eta \sqrt{\vec{p}_i^{\ 2} + m^2}  
\right)
+B(\vec{p}_i)\exp
\left(
i\eta \sqrt{\vec{p}_i^{\ 2} + m^2 }  
\right)
}{\sqrt{ (2\pi)^3 2 \sqrt{\vec{p}_i^{\ 2} +m^2 b^2}}
},\label{1}\\
g^{out}_{\vec{p}_f} &=& 
\Theta(\eta)
\frac{
\exp
\left(
-i\eta \sqrt{\vec{p}_f^{\ 2} + m^2}  
\right)
}{\sqrt{ (2\pi)^3 2 \sqrt{\vec{p}_f^{\ 2} +m^2 }}
}
\nonumber\\
&&+\Theta(-\eta)
\frac{
A(\vec{p}_f)\exp
\left(
-i\eta \sqrt{\vec{p}_f^{\ 2} + m^2 b^2}  
\right)
+B(\vec{p}_f)\exp
\left(
i\eta \sqrt{\vec{p}_f^{\ 2} + m^2 b^2}  
\right)
}{\sqrt{ (2\pi)^3 2 \sqrt{\vec{p}_f^{\ 2} +m^2 }}
},\nonumber\\
&&\ \label{2}
\end{eqnarray}
where 
\begin{eqnarray}
A(\vec{p}) &=& \frac{1}{2}\left(1+\sqrt{ \frac{\vec{p}^{\ 2} +m^2 b^2}
{\vec{p}^{\ 2} + m^2} }\right),\nonumber\\
B(\vec{p}) &=& \frac{1}{2}\left(1-\sqrt{ \frac{\vec{p}^{\ 2} +m^2 b^2}
{\vec{p}^{\ 2} + m^2} }\right).\nonumber
\end{eqnarray}
Substituting eqn(\ref{1}) and eqn(\ref{2}) into eqn(\ref{100}) and
taking the high momentum limit, $|\vec{p}_i|\rightarrow \infty$, it is
 shown that the
terms proportinal to $B$
 do not contribute to $W$ because of the  damping behavior of $B$.
Notice that taking the high momentum limit, the energy conservation law 
almost restores in the  following sense.
\begin{equation} 
|\vec{p}_i|\sim |\vec{k}|+|\vec{p}_i -\vec{k}|. \label{103}
\end{equation}
This can be read easily from the $\eta$ integration part in eqn
(\ref{100}). 
Contribution from the region where eqn(\ref{103}) does not hold  is
 severely  suppressed by the energy conservation factor. 
Using the polar 
coordinate decomposition 
$\vec{p}_i \cdot \vec{k} =|\vec{p}_i |k \cos \theta$ with $k=|\vec{k}|$
 and taking 
 $|\vec{p}_i|$ much larger than $m$,
it is easily derived that  
 no contribution comes from 
the $k$ integral region lying between $|\vec{p}_i|$and $\infty$. 
Hence we get 
\begin{eqnarray}
&&W(|\vec{p}_i|\sim \infty)\nonumber\\ 
&&= \frac{e^2 |\vec{p}_i|}
{2 (2\pi)^2} \int^{|\vec{p}_i|}_{|\vec{p}_i| \delta} dk 
\frac{k}{|\vec{p}_i| -k} \int^{\pi}_0  
d\theta\sin^3 \theta 
\nonumber\\
&&\times
\left| 
\left[k-\sqrt{\vec{p}_i^{\ 2} +m^2 }  
+\sqrt{(|\vec{p}_i| -k)^2 
+2|\vec{p}_i|k(1-\cos\theta)
+m^2 }
  \right]^{-1} \right.\nonumber\\
&&\left. 
-\left[k-\sqrt{\vec{p}_i^{\ 2} +m^2 b^2 }  
+\sqrt{(|\vec{p}_i| -k)^2 
+2|\vec{p}_i|k(1-\cos\theta)
+m^2 b^2}
  \right]^{-1}
\right|^2 .\nonumber\\
&&\ \ \ \ \ \ \ \ \ \ \ \ \ \ \ \ \ \ \ \ \ \ \ \ \ \ \ \ \ \ \label{1112}
\end{eqnarray}
We need  infra-red 
 cutoff $|\vec{p}_i|\delta$  in eqn(\ref{1112}) 
due to the existance of massless
 photon. This infra-red divergence
 is well known one in flat spacetime quantum
field theories with massless particles
 and it should be cancelled by an infra-red divergence of the self energy
term \cite{BN}. The cutoff $\delta$ is physically 
determined by resolving power of 
 soft photon observation. 
The $\theta$ integration in eqn(\ref{1112}) 
 can be
 straightforwardly calculated. After performing this
integration and  taking the high momentum limit 
$|\vec{p}_i|\rightarrow\infty$,  the $k$ integration is simplified and 
we finally obtain 
\begin{eqnarray}
W(|\vec{p}_i|\rightarrow \infty)
= \frac{e^2}{4\pi^2} \left(
 \ln \frac{1}{\delta}+\delta -1 \right) 
\left(\frac{1+b^2}{1-b^2}\ln \frac{1}{b^2} -2\right). \label{56}
\end{eqnarray}
Note that we have taken the helicity sum in eqn(\ref{56}). 
Instead, it is also possible to
evaluate $W(b)$ independently 
 with a fixed photon helicity. For left and 
right handed helicity, each probability is the same, a half of $W$ in
eqn(\ref{56}).

Furthermore we can also obtain   
the analytic forms of $W(b)$ in the spinor 
QED for the case of  eqn(\ref{8}). Because both of 
the charged fermion and photon 
have degree of helicity freedom, 4 helicity contributions 
 must be considered separately. 
The probability in the high momentum limit 
for 1/2 helicity fermion decaying into 
 fermion with helicitity $h_{fermion}$  
and photon with helicity $h_{photon}$ is 
denoted by $W(1/2; h_{fermion} , h_{photon} )$ and is given for each case
 as follows.
\begin{eqnarray}
&&W(1/2 ; 1/2 ,1)
= \frac{e^2}{8\pi^2} \left(
 \ln \frac{1}{\delta} \right) 
\left(\frac{1+b^2}{1-b^2}\ln \frac{1}{b^2} -2\right).\label{12}\\
&&W(1/2 ; 1/2, -1)
= \frac{e^2}{8\pi^2} \left(
 \ln \frac{1}{\delta} -\frac{\delta^2}{2} +2 \delta -\frac{3}{2} \right) 
\left(\frac{1+b^2}{1-b^2}\ln \frac{1}{b^2} -2\right).\label{13}\\
&&W(1/2;-1/2,1)
= \frac{e^2}{8\pi^2}  
\left(1- \frac{b}{1-b^2}\ln \frac{1}{b^2} \right).\label{10}\\
&&W(1/2;-1/2, -1)
= 0.\label{11}
\end{eqnarray}
No infra-red cutoff $\delta$  appears  in eqn(\ref{10}) and
eqn(\ref{11})  
because  spinflip of the fermion enables observers to distinguish the
bremsstrahlung from the self energy process.

\bigskip

There exists a very useful aspect of $W$ in the high momentum limit.  
 It is supposed that the results eqn(\ref{56})$\sim$eqn(\ref{11}) are
exact not only for the special way of the expansion given by eqn(\ref{8}) 
but also arbitrary way  satisfying  eqn(\ref{1001}) $\sim$ eqn(\ref{1013}).
This implies  that 
$W(|\vec{p}_i|\rightarrow \infty)$ possesses a remarkable universality
 with respect to the ways of the cosmic expansion.
 This property may be explained as the
Lorentz contraction effect from the view point of the high energy
particle. Imagine a particle running in the comoving frame. 
Suppose that the Universe begins to expand when the particle passes
through  a point A and the Universe ceases to expand when the particle 
reaches point B. 
The particle catches energy 
 from the expansion only while running
from A to B. Taking the high momentum limit, 
the length between A and B contracts 
to zero in the rest frame of the particle.
 Therefore the particle cannot see the details of 
the way how the Universe expands and thus 
the universality of $W$ crops up.

To see the universality more quantitatively, we shall discuss  
 the scalar QED with an adiabatically 
 slow evolution of the scale factor $a(\eta)$ satisfying 
 eqn(\ref{1001}) $\sim$ eqn(\ref{1013}). 
 In the zeroth
order adiabatic approximation(WKB approximation) the mode functions 
 satisfying eqn(\ref{9}) is written as 
\begin{eqnarray}
g^{in}_{\vec{p}} \sim g^{out}_{\vec{p}} \sim 
\frac{
\exp \left[ i\vec{p}\cdot\vec{x} -i\int^\eta_0  d\eta' 
\sqrt{\vec{p}^{\ 2}+ m^2 a(\eta ')^2 } 
\right]
}
{
\sqrt{ (2\pi)^3 2\sqrt{\vec{p}^{\ 2} +m^2 a(\eta )^2} }
}.\label{101}
\end{eqnarray}
Substituting eqn(\ref{101}) into eqn(\ref{100}) and introducing the polar 
coordinate decomposition; 
$\vec{p}_i \cdot \vec{k} =|\vec{p}_i |k \cos \theta$,
we get
\begin{eqnarray}
&&W \sim
 \frac{e^2 |\vec{p}_i|^2}
{2 (2\pi)^2} \int^{\infty}_{|\vec{p}_i| \delta} dk k \int^{\pi}_0  
d\theta\sin^3 \theta 
\nonumber\\
&&\times
\left| \int^{\infty}_{-\infty}
d\eta \ 
\frac{
\exp \left[ik\eta -i\int^\eta d\eta 
'\sqrt{\vec{p}_i^{\ 2} +m^2 a(\eta ')^2}  
+i\int^\eta d\eta '\omega(\vec{p}_i -\vec{k},\eta ')  \right]
}
{
\sqrt{\omega(\vec{p}_i -\vec{k},\eta ) \sqrt{\vec{p}_i^{\ 2} +m^2 a(\eta 
)^2}
} }
\right|^2\nonumber\\
&&\ \ \ \ \ \ \ \ \ \ \ \ \ \ \ \ \ \ \ \ \ \ \ \ \ \ \ \ \ \ \label{102}
\end{eqnarray}
where $\omega(\vec{p}_i -\vec{k},\eta )=
\sqrt{(|\vec{p}_i| -k)^2 
+2|\vec{p}_i|k(1-\cos\theta)
+m^2 a^2}$ and $|\vec{p}_i|\delta$ is the inra-red cutoff. 
 Consider the high momentum 
limit in eqn(\ref{102}). As mentioned before nonvanishing contribution to $W$ 
comes from the integral region where the momentum holds 
the relation eqn(\ref{103}). 
 Thus it is enough to restrict the momentum region  
 between $|\vec{p}_i|$ and 
 $|\vec{p}_i|\delta$.
 Here it should be searched 
which  integral region of $\theta$ 
contributes, accompanied with the  influence of  $a(\eta)$, 
to the nonvanishing value of $W$ in
 eqn(\ref{102}).
 Due to eqn(\ref{103}), only emittion to nearly forward direction
 ($\theta\sim0$)
 is permitted
and  especially the  integral region of $\theta$ safisfying  
$$0\leq \theta\leq O\left(m/|\vec{p}_i|\right)$$
 gives  the scale factor dependence to the $W$.
 Several  expansions like 
\begin{eqnarray}
\sqrt{\vec{p}_i^{\ 2} +m^2 a(\eta ')^2} 
\sim |\vec{p}_i |+\frac{m^2 a(\eta ')^2}{2|\vec{p}_i |} \nonumber
\end{eqnarray}
 yield finally 
\begin{eqnarray}
&&W \sim
 \frac{e^2 |\vec{p}_i|}{2 (2\pi)^2}
\int^{|\vec{p}_i|}_{|\vec{p}_i|\delta}
 dk 
\frac{k}{|\vec{p}_i| -k} 
\int^{O\left(\frac{m}{|\vec{p}_i|}\right) }_0
d\theta\ \theta^3 
\nonumber\\
&&\times
\left|
 \int^{\infty}_{-\infty}
d\eta \ 
\exp \left[
-i\int^\eta_0 d\eta ' 
\left(
\frac{m^2 a(\eta ')^2 }{2|\vec{p}_i|}
-\frac{m^2 a(\eta ')^2}{2(|\vec{p}_i| -k)} 
-\frac{|\vec{p}_i|k\theta^2 )}{2(|\vec{p}_i| -k)}
\right)
\right]
\right|^2 \nonumber\\
&&=
\frac{ e^2 }{(2\pi)^2} \int^1_{\delta} dy \frac{y}{1-y}
\int^{O(1)}_0 dz z^3 \nonumber\\
&&\times
\left|
 \int^{\infty}_{-\infty}
d\tilde{\eta} \ 
\exp \left[
-i\int^{\tilde{\eta}}_0 d\tilde{\eta}' 
\left(
C\left(\frac{2\lambda|\vec{p}_i|}{m^2}\tilde{\eta}'\right)^2 
-\frac{C(\frac{2\lambda|\vec{p}_i|}{m^2}\tilde{\eta}')^2 
}{1-y} 
-\frac{y z^2}{1-y}
\right)
\right]
\right|^2 
,\nonumber\\
&&\ \ \ \ \ \ \ \ \ \ \ \ \ \ \ \ \ \ \ \ \ \ \ \ \ \ \ \ \ \ \ \ \label{104}
\end{eqnarray}
where we change the integral variables in the following way,
\begin{eqnarray}
k&=& |\vec{p}_i| y,\nonumber\\
\theta&=& \frac{m}{|\vec{p}_i|}z,\nonumber\\
\eta&=& \frac{2|\vec{p}_i|}{m^2} \tilde{\eta}.\nonumber
\end{eqnarray}
Note that the function 
$C(\frac{2\lambda|\vec{p}_i|}{m^2}\tilde{\eta})$ in eqn(\ref{104})
 approaches in the high momentum limit  to a step function,
\begin{eqnarray}
C\left(\frac{2\lambda|\vec{p}_i|}{m^2}\tilde{\eta}\right) \sim
\Theta (\tilde{\eta}) + b\ \Theta(-\tilde{\eta}).\label{1.11}
\end{eqnarray} 
Therefore the value of $W$ for arbitrary adiabatical cosmic 
expansion satisfying  eqn(\ref{1001}) $\sim$ eqn(\ref{1013})
 must equal to  the specified value for  
 eqn(\ref{8}), and the universality is surely realized.
 If we dismiss the adiabatic approximation, 
the mode functions have reflection wave 
terms like in eqn(\ref{1}) and eqn(\ref{2}). However amplitude of the 
reflection waves vanishes in the high momentum limit and the
universality are thought to survive. 

\bigskip
Here we have a comment on the rate of the
geometric bremsstrahlung.  
Comparing with the classical results in the section 2, 
it is noticed from  
 eqn(\ref{56})$\sim$ eqn(\ref{11}) 
that 
 interaction rate of quantum geometric bremsstrahlung
 does not so large compared with classical one. 
This is due to the fact that quantum effect smears position of
the classical point particle and dilutes the charge density.

\bigskip
Now 
it is worth considering implication of the results;
 eqn(\ref{56})$\sim$eqn(\ref{11}) in connection with 
 the conformal symmetry. 
 Since the massless limit $m\rightarrow 0$  
 forces the speed of the  particle to reach the light velocity, 
the universality with respect to the way of the cosmic 
expansion is  maintained.
 In the lowest pertubation of
 the QED,
the massless limit  is shown to be equivalent with  
 the $|\vec{p}_i| \rightarrow \infty$ limit.
 Therefore again the same results, eqn(\ref{56})$\sim$
eqn(\ref{11}), come up in the $m\rightarrow 0$ limit and  $W$ really 
possesses the non-vanishing value.    One might naively 
expect for the massless case  
that the amplitude in the conformally flat spacetime vanishes
 as  in the Minkowskian spacetime, by virtue of the conformal symmetry
 guaranteed at least classical mechanically.
 However this is {\it not} 
true unless $b=1$ as argued above.

\bigskip

Next let us discuss the rate of the geometric bremsstrahlung.  
Taking large expansion limit $b\sim 0$,
 the transition probablity can be typically expressed  
 as 
\begin{eqnarray}
W(b\sim 0) \sim O(1) \frac{N^{\ast}e^2}{4\pi^2} \ln \frac{1}{b } ,
\end{eqnarray}
where $N^{\ast}$ is the number of final modes. In our spinor QED model 
 we take $N^{\ast} =2$ due to contributions of
 eqns (\ref{12}) and (\ref{13}). However in the standard model  
 we have more particles and $N^{\ast} \sim O(10)$.  
Moreover extended theories (like GUT and SUSY) can gives us
 $N^{\ast} \sim O(100)$. 
 
The process is thought to occur when $W\sim 1$.  Thus 
when the Universe expands enough to satisfy
\begin{eqnarray}
\frac{1}{b}=\frac{a_f}{a_i}\sim 
\exp \left[ \frac{4\pi^2 O(1)}{N^{\ast} e^2} \right] \label{555}
\end{eqnarray}
 the event will take place. If we specify a model of the Unvierse
evolution, we can get the rate itself. For example 
let us assume  that the expansion is dominated by 
the radiation, $a(t) \propto t^{1/2}$. Then we get the following 
 estimation for the transition rate.
\begin{eqnarray}
\Gamma \sim 
\frac{1}{t_f} 
= b^2 \frac{1}{t_i}
\sim \exp\left[ -  \frac{8\pi^2 O(1)}{N^{\ast} e^2} \right] H_i
\sim e^{-\frac{O(100)}{N^{\ast}} }H_i,
\end{eqnarray}
where $H_i =1/t_i$ is the Hubble parameter at the initial time.
 Thus we cannot neglect the quantum bremsstrahlung 
 naively when $N^{\ast}\sim O(100)$. It might be also worth 
 reminding that  rates of ohther gauge interection
processes different from the geometric bremsstrahlung is much smaller
than the Hubble parameter when the temparature of the
Universe is higher than $O(10^{15})$ GeV. Thus the bremsstrahlung may 
 be the most dominant process in such a  early era.

\section*{ Acknowledgement}

The authors wish to thank T. Goto, I.Joichi, 
T. Moroi, M. Tanaka and M. Yoshimura for 
 fruitful discussions. We also thank to K.Hikasa and J.Arafune for 
their critical comments.

\end{document}